\documentclass{nature_arx}
\usepackage[format=plain, font=bf, justification=justified, singlelinecheck=false]{caption}
\usepackage{graphicx}

\newcommand*{\afe}{[$\alpha$/Fe]}
\newcommand*{\alp}{$\alpha$}
\usepackage{amsmath,lipsum}
\newcommand{\mpm}{\mathbin{\smash{%
\raisebox{0.35ex}{%
            $\underset{\raisebox{0.4ex}{$\smash -$}}{\smash+}$%
            }%
        }%
    }%
}
\makeatletter
\newcommand{\mypm}{\mathbin{\mathpalette\@mypm\relax}}
\newcommand{\@mypm}[2]{\ooalign{%
  \raisebox{.1\height}{$#1+$}\cr
  \smash{\raisebox{-.6\height}{$#1-$}}\cr}}
\makeatother
\title{Evidence for the accretion origin of halo stars with an extreme r-process enhancement}

\author{Qian-Fan Xing$^{1}$, Gang Zhao$^{1}$, Wako Aoki$^{2,3}$, Satoshi Honda$^{4}$, Hai-Ning Li$^{1}$, Miho N. Ishigaki$^{5}$, Tadafumi Matsuno$^{2,3}$}

\begin{document}
\pagenumbering{gobble}

\maketitle

\begin{affiliations}
 \item Key Laboratory of Optical Astronomy, National Astronomical Observatories, Chinese Academy of Sciences, Beijing 100101, China; gzhao@nao.cas.cn, qfxing@nao.cas.cn
 \item National Astronomical Observatory of Japan, 2-21-1 Osawa, Mitaka, Tokyo 181-8588, Japan
  \item Department of Astronomical Science, School of Physical Sciences, The Graduate University
of Advanced Studies (SOKENDAI), 2-21-1 Osawa, Mitaka, Tokyo 181-8588, Japan
 \item Center for Astronomy, University of Hyogo, 407-2, Nishigaichi, Sayo-cho, Sayo, Hyogo 679-5313, Japan
 \item Kavli Institute for the Physics and Mathematics of the Universe, The University of Tokyo, Kashiwa, Chiba 277-8583, Japan
\end{affiliations}

\begin{abstract}

Small stellar systems like dwarf galaxies are suggested to be the main building blocks of our
Galaxy by numerical simulations\cite{Bullock_2005} in
Lambda CDM models. The existence of star streams like Sagittarius
tidal stream\cite{Ibata_1994} indicates that dwarf galaxies play a role
in the formation of the Milky Way. However, it is unclear how many and
what kind of stars in our Galaxy are originated from satellite dwarf
galaxies, which could be constrained by chemical abundances of
metal-poor stars. Here we report on
the discovery of a metal-poor star with an extreme r-process
enhancement and \alp-element deficiency. In this star, the abundance ratio of the
r-process element Eu with respect to Fe is more than one order of magnitude
higher than the Sun and the metallicity is $\left. 1 \middle/ 20 \right.$ of the solar one.
Such kind of stars have been found in present-day dwarf galaxies, providing
the clearest chemical signature of past accretion events.
The long timescale of chemical evolution of the host dwarf galaxy expected from the abundance
of \alp~element with respect to Fe suggests that the accretion occurred in a relatively late phase
compared to most of the accretions that formed the bulk of the Milky Way halo.

\end{abstract}

Chemical abundance ratios of stars reflect the contributions of nucleosynthesis by earlier generation
stars that have provided the products through stellar mass-loss and supernova explosions.
The products are dependent on progenitor masses. High-mass ($>$$8\,{\mathrm M}_{\odot}$) and
short-lived stars dominate the chemical enrichment at the early stage of the
Galaxy\cite{Tinsley_1979}. The yields of type II supernova (SN II)
resulting from explosions of massive stars are rich in
\alp-elements (e.g. Mg, Si, Ca, and Ti), resulting in
\alp-element enhancements ([Mg/Fe]$\sim$$+0.4$) of the Galactic
metal-poor stars\cite{McWilliam_1997}. There is, however, evidence of
scatter in \afe~in Galactic halo stars, and a small number
of metal-poor stars show particularly low [Mg/Fe] values (\alp-poor stars)
\cite{Ivans_2003,Caffau_2013}, implying that the chemical enrichment history is different from that for the
bulk of halo stars (see Figure 1a).

Many of metal-poor stars with low [Mg/Fe] have been
found in dwarf galaxies around the Milky
Way\cite{Tolstoy_2009}. Figure 2a shows the Mg abundance
distribution of Fornax, Sculptor, Draco, Carina and Sextans dwarf
galaxies, in comparison with those of Galactic halo stars. These dwarf galaxies
are called classical dwarf galaxies and have luminosity larger than $10^{5}$ solar
luminosity, whereas galaxies having smaller luminosity are called
ultra-faint dwarf galaxies. While all of these classical dwarf galaxies
possess Mg-enhanced stellar population similar to stars in the
Galactic halo in particular at very low metallicity, the declines of their [Mg/Fe] occur at much lower
metallicities and decrease to lower values than in the Milky Way\cite{Letarte_2010}. Contributions from type Ia supernova
(SN Ia) are the most promising explanation for the low [Mg/Fe] of these dwarf galaxies,
although variations of the initial mass function are also argued\cite{Kobayashi_2014}. Since the time-scale of SN Ia is
longer than that of core-collapse supernovae, which yields high \afe~ratios, \alp-poor stars should have been formed in environment with
slower chemical evolution as expected for dwarf galaxies.
The \alp-poor stars in
the Galactic halo could provide direct link
between halo building blocks and satellite dwarf galaxies, but
evidence for accretion from such small stellar systems in chemical
abundance ratios in the halo stars is still limited.

LAMOST J112456.61+453531.3 (hereafter J1124+4535) was
identified as a candidate of \alp-poor star\cite{Xing_2014, Xing_2015} in the
Galactic halo by low-resolution spectroscopy in the Large Sky Area Multi-Object Fiber Spectroscopic
Telescope (LAMOST) survey\cite{Zhao_2012}.A high-resolution
(R $= 45,000$) follow-up spectrum for J1124+4535 was obtained with the
Subaru Telescope High Dispersion Spectrograph\cite{Noguchi_2002} (HDS) in February
2017, covering 4030-6800 \AA~with S/N = 60 and 30 at 5200 \AA~and 4130 \AA, respectively. An effective temperature of 5180 K, a surface gravity
of log g $= 2.7$ and a micro-turbulent velocity of 1.5 km s$^{-1}$ are determined by spectroscopic analysis of Fe lines (see Methods).
The chemical abundances for a total of 24 elements, as well as upper limits for two others, are determined
by the standard abundance analysis and spectrum synthesis techniques based on model atmospheres\cite{Castelli_1997} (see Table 1).
J1124+4535 has [Mg/Fe] $=-0.31$ (here
[A/B]=$\log_{10}({N_{A}}/{N_{B}})_{*} -
\log_{10}({N_{A}}/{N_{B}})_{\odot}$ for elements A and B) with [Fe/H] of $-1.27$, confirming
the Mg deficiency discovered from the low-resolution spectrum.  The
[Mg/Fe] of J1124+4535 is more than $0.6$ dex below the typical [Mg/Fe]
of the Galactic stars with similar metallicity (see Figure 1a).
This significant departure from the general Mg enhancement trend of the Galactic
metal-poor stars implies J1124+4535 is formed in a separate system with
relatively low star formation rate (SFR) such as dwarf galaxies.

A more remarkable feature of J1124+4535 is the extreme enhancement of
r-process elements. Figures 1b and 1c show the
[Eu/Fe] and [Eu/Mg] of J1124+4535 in comparison with
Galactic stars from the literature.
As shown in Figure 1d, the abundances of heavy neutron-capture elements
of J1124+4535 are in agreement with the scaled solar-system r-process
pattern\cite{Arlandini_1999}, consistent with the heavy neutron-capture element (Z $\geq$ $56$) abundance pattern
of r-process enhanced stars in the halo and dwarf galaxies\cite{Hansen_2017}.
Most of r-II stars\cite{Hansen_2017} (r-process-enhanced stars with [Eu/Fe]$>$$+1$)
are found in the Galactic halo at extremely low metallicity ([Fe/H]$<$$-2.5$), and
the fraction is quite low: less than $5\%$ stars in this metallicity
range\cite{Hansen_2017}. These extremely metal-poor stars would be formed from gas
clouds that were affected by events that yielded large amount of
r-process elements (most likely mergers of binary neutron stars).
J1124+4535 is a unique object that shows large enhancement of
r-process elements like r-II stars but is just moderately metal-poor.
Such large excesses
of r-process elements are not expected in stars with higher
metallicity (moderately metal-poor stars), because the chemical
composition of these stars are usually determined by contributions of a large
number of nucleosynthesis events until the stars were formed, and any
single event could not significantly change the abundance ratios.
Only a few of such moderately metal-poor r-II stars have been
found in the Galactic halo (HD\,222925 and J1802-4404)\cite{Roederer_2018, Hansen_2018}
and bulge\cite{Howes_2016}. J1124+4535 is unique among them for being deficient in
\alp-elements.

Interestingly, however, such moderately metal-poor stars with large
enhancement of r-process elements are found in dwarf galaxies around
the Milky Way (see Figure 2b).  For instance, COS\,82 in Ursa Minor (UMi) dwarf galaxy has
[Fe/H] $=-1.4$ and [Eu/Fe] $=+1.2$ with r-process abundance pattern\cite{Aoki_2007}.
The extremely large enhancement of
r-process elements of J1124+4535 with [Fe/H] $=$$-1.27$ and
[Eu/Fe] $=+1.1$ is similar to this star in UMi dwarf galaxy. Moreover, the sub-solar
[Mg/Fe] of J1124+4535 agrees well with the trend in the UMi stars (see Figure 2a).
Figure 3 shows the abundances of J1124+4535 and UMi stars in comparison with Galactic
halo stars. Generally, the abundances of J1124+4535 are in a good agreement with those
of UMi stars. They exhibit abnormally low Carbon abundances. Their Na, Sc, Ni and Zn abundances are clearly lower than that of Galactic halo stars.
The UMi stars exhibit \alp-element (Mg, Si, Ca and Ti) abundances ratios consistent with those of halo stars
at very low metallicity, but their abundance ratios steadily decrease to sub-solar values with
increasing metallicity. The [Cr/Fe] and [Mn/Fe] of UMi stars appear to behave as halo stars and increase
with increasing [Fe/H]. The [Sr/Fe] values of J1124+4535 and UMi stars with [Fe/H] $>$$-2$ are lower
than typical values of halo stars, in contrast to large enhancement of heavier neutron-capture elements in
J1124+4535 and UMi COS\,82. This implies that these stars are not significantly affected by s-process, in particular
weak s-process, nor by weak r-process\cite{Travaglio_2004}.
The overall abundance
pattern of J1124+4535 suggests it has been formed in a dwarf galaxy similar to UMi.
This star would provide the strongest evidence for accretion of
dwarf spheroidal (dSph) galaxies from stellar chemical composition obtained so far.

The recent discovery of the r-process-enhancement in extremely
metal-poor stars in the ultra-faint dwarf galaxy Reticulum
II\cite{Ji_2016} suggests that r-II stars in the Galactic halo are
formed in small stellar systems affected by an r-process event. Recent
chemical evolution models for dwarf galaxies support this
interpretation\cite{Hirai_2017}. The first detection of gravitational-wave from
a neutron star merger (NSM) confirms the contribution of r-process elements
from the NSM\cite{Kasen_2017}.  The long time-scale of the orbital
decay of a neutron star binary results in delayed enrichment from the
NSM. In order to accommodate such delay, the extremely r-process-enhanced
stars originated from NSM with [Fe/H]$\sim$$-3$ must have
been formed in environment with slow chemical evolution. Because of lower
star formation efficiencies, [Fe/H] of interstellar medium in dwarf
galaxies would be lower than in the Milky Way when early NSM occurred.
Hence, r-process-enhanced stars could be formed at extremely low
metallicity in small stellar systems even if the time-scale of mergers
of binary neutron stars is assumed to be long.

The r-process-enhanced star J1124+4535 would also be originated from a
dwarf galaxy. The situation is, however, different from r-II stars with extremely low metallicity
previously known. The metallicity of this object is more than one order
of magnitude higher. The amount of Eu in this object
([Eu/H] $=-0.17$) is one order of magnitude higher than in extremely metal-poor
r-II stars ([Eu/H]$\sim$$-1.3$). The amount of Eu in Reticulum II (Ret II) stars could
be explained by pollution of a 10$^{6}$\,M$_{\odot}$ cloud\cite{Ji_2016} by a single
r-process event that yields Eu $10^{-4.3}$\,M$_{\odot}$. If the same
yields of the r-process event is assumed, the gas cloud of the
progenitor of J1124+4535, as well as moderately metal-poor stars with
r-process excess in dwarf galaxies, should be as small as
$10^{5}$\,M$_{\odot}$. It would be a challenge to model efficient capture of the energetic
ejecta from NSM in such low-mass gas clouds. It is noted that the amount of Eu in Ret II could also
be explained by magnetorotationally driven supernovae (also called as Jet-SN)\cite{Nishimura_2015}. These particular
supernovae synthesize large quantities of r-process material that are similar to the r-process yields of NSM, resulting
in the difficulty in distinguishing these two sites based on the current data\cite{Ji_2016}. An alternative possibility is that the progenitor
dwarf galaxy of J1124+4535, as well as Ursa Minor dwarf galaxy that includes COS\,82,
has been enriched by more than one r-process event. In that case, however, there is
a difficulty in explaining the fact that only a small number of stars are exceptionally enhanced
in r-process elements.

Similar estimations on ejecta and gas cloud masses ($10^{3}$ to $10^{4}$\,M$_{\odot}$) were made for another moderately
metal-poor r-II star, HD\,222925 ([Eu/Fe] $=+1.33$, [Fe/H] $=-1.47$ and \afe~$=+0.4$), recently
discovered\cite{Roederer_2018}. In contrast to J1124+4535, the \afe~ratio of HD\,222925 is as high as those
of typical halo stars. This indicates that HD\,222925 was formed in a system with relatively rapid chemical
evolution, whereas J1124+4535 was formed in a galaxy with slower chemical evolution of longer duration. Hence,
J1124+4535 is a unique object that provides a clear signature of accretion of evolved dwarf galaxy, like current
dwarf galaxies such as Ursa Minor, to the Milky Way halo in a later phase.

It is not easy to identify the record of accretion of dSph galaxies in chemical abundances of individual stars. The deficiency of \alp~element and large excess of r-process elements found in J1124+4535 provide clear evidence for accretion of a small galaxy that is similar to present-day dwarf galaxies. The \alp-element deficiency of metal-poor stars is regarded as a result of relatively slow chemical evolution of their birth site. The origin of r-process-excess in moderately metal-poor stars may caused by a single r-process event in a protogalactic fragment with small mass. Stars similar to J1124+4535 are very rare and have only been found in dSph galaxies. The discovery of J1124+4535 could be a start of chemical identification of stars accreted from dSph galaxies.

\begin{table}
%\footnotesize
\scriptsize
\caption{Chemical abundances of J1124+4535}
\label{table:abunds}
\sffamily
\begin{tabular}{lcrcc}
\hline
Element X & log(${N_{X}}/{N_{H}}$+ 12) & [X/Fe] & $N_{\rm lines}$ & $\sigma$ \\
\hline
Fe \scriptsize{I} & 6.23 & 0.00 & 95 & 0.10  \\
Fe \scriptsize{II} & 6.23 & 0.00 & 10 & 0.09 \\
C (CH) & 6.74 & $-$0.42 & $\cdots$ & 0.20 \\
Na \scriptsize{I} & 4.08 & $-$0.89 & 2 & 0.13 \\
Mg \scriptsize{I} & 6.02 & $-$0.31 &  3 & 0.09 \\
Si \scriptsize{I} & 6.06 & $-$0.18 & 4 & 0.08 \\
Ca \scriptsize{I} & 4.98 & $-$0.09 & 12 & 0.08 \\
Sc \scriptsize{II} & 1.73 & $-$0.15 & 6 & 0.09 \\
Ti \scriptsize{I} & 3.63 & $-$0.05 & 8 & 0.09 \\
Ti \scriptsize{II} & 3.81 & 0.13 & 7 & 0.10 \\
Cr \scriptsize{I} & 4.04 & $-$0.33 & 6 & 0.08 \\
Mn \scriptsize{I} & 3.84 & $-$0.32 & 4 & 0.10 \\
Ni \scriptsize{I} & 4.60 & $-$0.35 & 6 & 0.10 \\
Zn \scriptsize{I} & 2.92 & $-$0.37 & 2 & 0.09 \\
Sr \scriptsize{II} & 1.24 & $-$0.36 & 2 & 0.10 \\
Y \scriptsize{II} & 0.76 & $-$0.18 & 4 & 0.13 \\
Zr \scriptsize{II} & 1.44 & 0.13 & 2 & 0.14 \\
Ba \scriptsize{II} & 1.15 & 0.24 & 4 & 0.12 \\
La \scriptsize{II} & 0.51 & 0.68 & 6 & 0.09 \\
Ce \scriptsize{II} & 0.72 & 0.41 & 4 & 0.13 \\
Pr \scriptsize{II} & 0.20 & 0.75 & 4 & 0.13 \\
Nd \scriptsize{II}  & 0.81 & 0.66 & 15 & 0.14 \\
Sm \scriptsize{II} & 0.52 & 0.83 & 14 & 0.13 \\
Eu \scriptsize{II} & 0.35 & 1.10 & 1 & 0.10 \\
Gd \scriptsize{II} & 0.63 & 0.83 & 2 & 0.13 \\
Dy \scriptsize{II} & 0.74 & 0.91 & 2 & 0.15 \\
Er \scriptsize{II} & $<$ 0.7 & $<$ 1.05 & 1 & $\cdots$ \\
Th \scriptsize{II} & $<-$0.16 & $<$ 1.09 & 1 & $\cdots$ \\
\hline
\end{tabular}
\rmfamily
\\
\\ Abundance ratios for J1124+4535 as derived from Subaru/HDS spectra. The abundance of Fe I and solar abundances from literature\cite{Asplund_2009} are
adopted for obtaining [X/Fe].
The error is the quadratic sum of the random error and errors due to the stellar parameter uncertainties ($\delta$$T_{\rm eff}$ = 100 K, $\delta$log g = 0.3 dex,
and $\delta$$V_{\rm micro}$ = 0.3 km s$^{-1}$). The random error is estimated to be $\sigma$$_{\upsilon}$$N$$_{\rm lines}^{-1/2}$, where $\sigma$$_{\upsilon}$ is the dispersion around the mean abundance obtained from different lines for each species, and $N$$_{\rm lines}$ is the number of lines used to derive abundance. The $\sigma$$_{\upsilon}$ of Fe I is adopted for species with $N$$_{\rm lines}$ $<$ 3. For C and Eu abundances measured by spectral synthesis, the errors are estimated based on comparison with synthetic spectra.
\\
\\
\end{table}

\begin{figure}
\centering
\includegraphics[width=0.95\columnwidth]{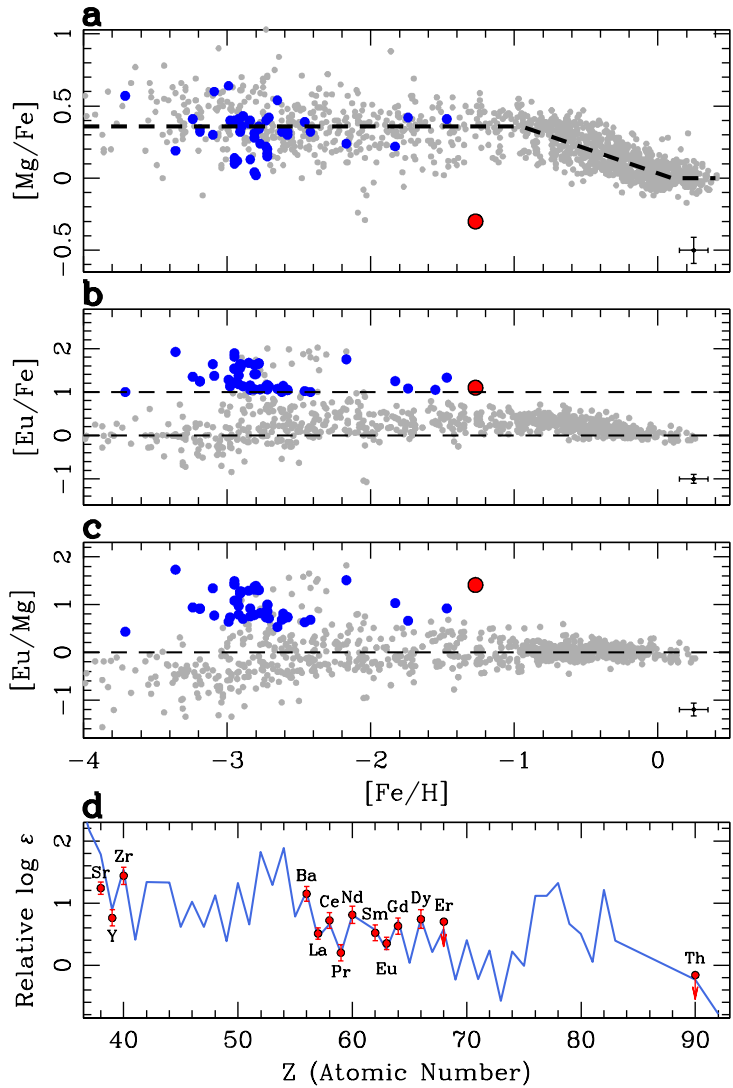}
\captionsetup{font=footnotesize}
\caption*{\textbf{Figure 1:}
\textbf{Abundance ratios of Mg, Eu as a function of metallicity in J1124+4535 and Galactic halo and disk stars.} \textbf{a:} [Mg/Fe] versus [Fe/H] for J1124+4535 (red filled circle), Galactic halo and disk stars (gray filled circles\cite{Hansen_2017,Frebel_2010,Ishigaki_2013}). The blue filled circles represent the known r-II field stars\cite{Hansen_2017,Frebel_2010,Roederer_2018b}, including moderately metal-poor r-II stars HD\,222925 ([Eu/Fe] $=+1.33$, [Fe/H] $=-1.47$ and [Mg/Fe] $=+0.4$) and J1802-4404 ([Eu/Fe] $=+1.05$ and [Fe/H] $=-1.55$). The dashed line indicates the trend found in the Milky Way field stars. \textbf{b:} Same as Fig. 1a, but for [Eu/Fe]. \textbf{c:} Same as Fig. 1a, but for [Eu/Mg]. \textbf{d:} Abundances of neutron-capture elements in J1124+4535 (red dots) compared with the r-process component in solar system (solid line)\cite{Arlandini_1999}, scaled to match the abundances of Ba to Dy in J1124+4535. The $\sigma$ of derived abundances are shown as error bars.}
\end{figure}

\begin{figure}
\centering
\includegraphics[width=0.95\columnwidth]{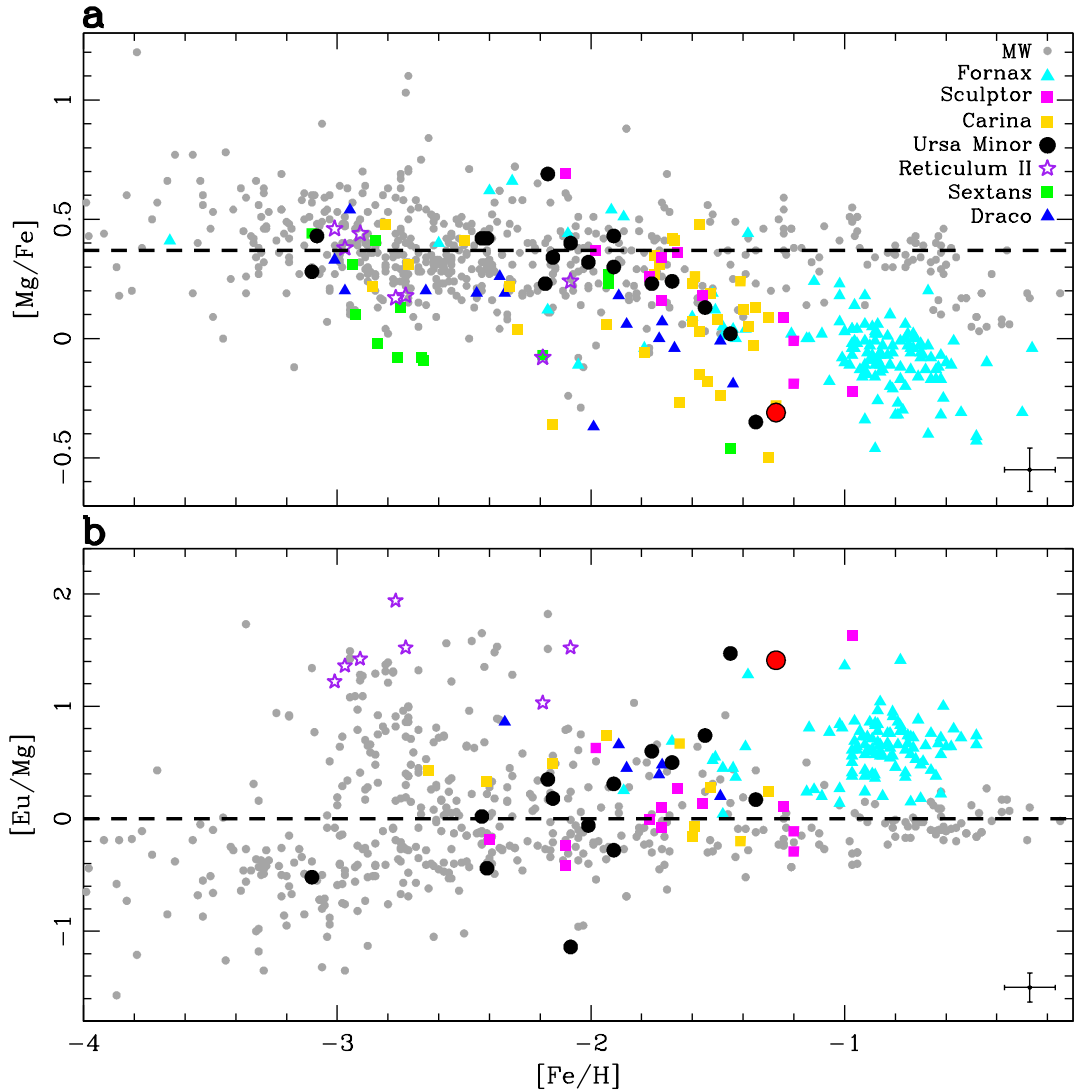}
\captionsetup{font=footnotesize}
\caption*{\textbf{Figure 2:}
\textbf{Abundance ratios of Mg, Eu as a function of metallicity in J1124+4535 and other metal-poor stars.}
\textbf{a:} [Mg/Fe] versus [Fe/H] for J1124+4535 (red filled circle), Galactic halo field stars (gray filled circles\cite{Frebel_2010,Ishigaki_2013}), stars in the ultra-faint dwarf galaxy Reticulum II\cite{Ji_2016} and stars in classical dwarf galaxies\cite{Suda_2017} named Fornax, Sculptor, Carina, Ursa Minor, Sextans and Draco. The dashed line represents the average value of [Mg/Fe] in the Galactic halo.
\textbf{b:} Same as Fig. 2a, but for [Eu/Mg].}
\end{figure}

\begin{figure}
\centering
\includegraphics[width=0.95\columnwidth]{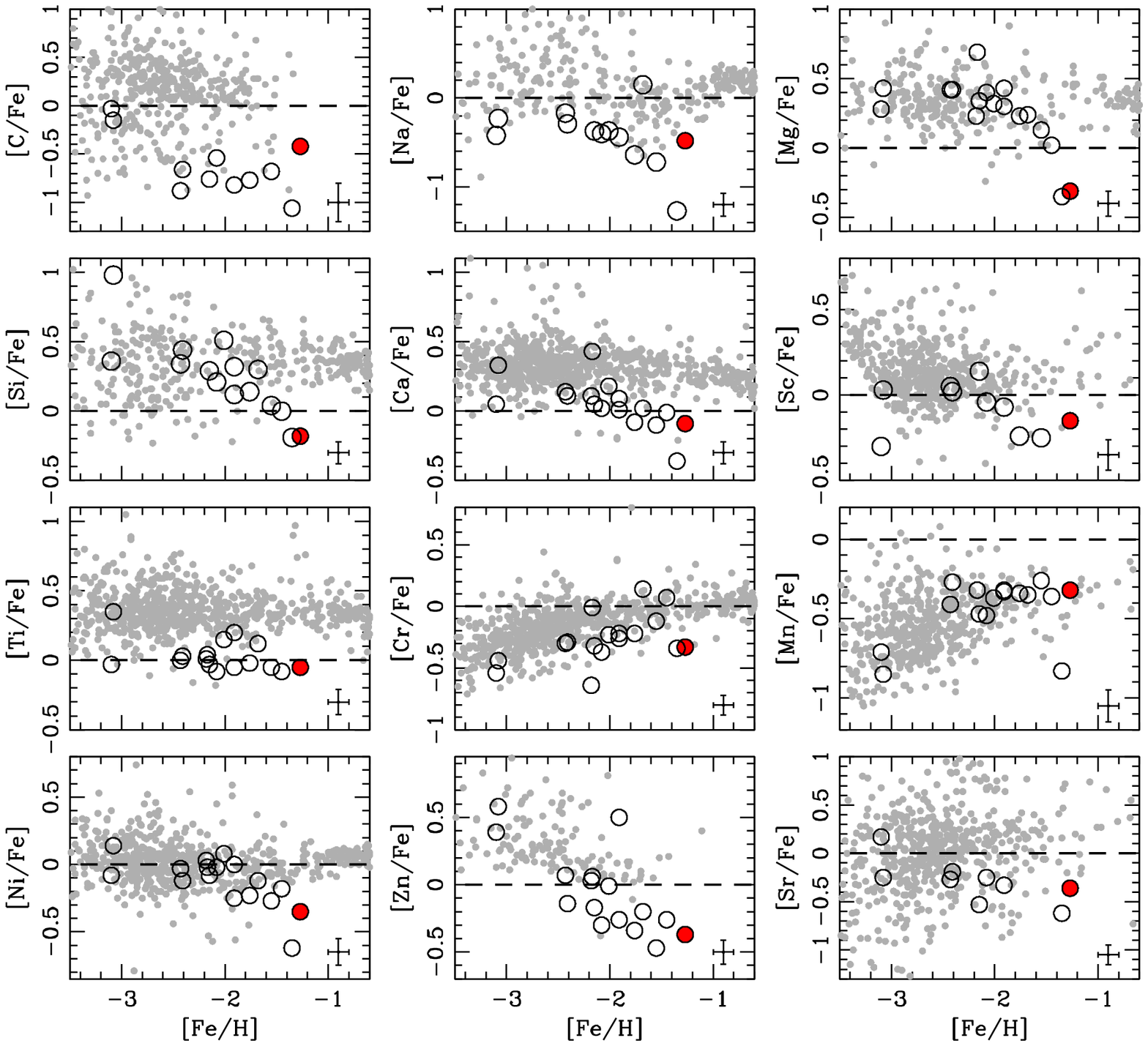}
\captionsetup{font=footnotesize}
\caption*{\textbf{Figure 3:}
\textbf{Abundance ratios as a function of metallicity in J1124+4535 and stars in the classical dwarf galaxy UMi.} In twelve elements, J1124+4535 (red filled circle) is compared with the sample of UMi stars (black open circles\cite{Suda_2017}) and halo stars (gray filled circles\cite{Frebel_2010,Ishigaki_2013})
}
\end{figure}

%----------------------------------------------------------------------------------------
%   Main References
%----------------------------------------------------------------------------------------

\newpage

\begin{methods}
\textbf{Discovery.} LAMOST J1124+4535 (apparent magnitude V = 13.98) is located at RA = 11 h 24 min 56.7 s, dec. = +45$^{\circ}$35$'$ 32$''$ (equinox 2000). The low-resolution (R$\sim$1800) spectrum was obtained by LAMOST on 8 January 2015. The [Mg/Fe] of the observed spectrum was determined by fitting the Mg Ib line. J1124+4535 was selected as a candidate of \alp-poor star according to its sub-solar [Mg/Fe].\\    \textbf{High-resolution spectroscopy.} The high-resolution spectra of LAMOST J1124+4535 were obtained on 16 February 2017 with Subaru/HDS. The observation was taken with a spectral resolution of R = 45,000. Data reduction was carried out with the IRAF package.\\   \textbf{Stellar parameter and abundance determination.} Stellar parameters were determined based on high-resolution spectra. Effective temperature was derived by demanding that the abundances of individual Fe I absorption lines are independent of their excitation potentials. Microturbulent velocity was determined from the atomic Fe I absorption lines by ensuring that the derived abundances exhibit no trend with the reduced equivalent widths. Surface gravity was determined from a reasonable agreement between two ionization stages of Fe (Fe I and Fe II). The Fe abundances derived from 95 Fe I lines and 10 Fe II lines are used for the analysis. Although the uncertainty of the parallax provided by Gaia DR2\cite{Lindegren_2018} is large (relative error of 40\%), we derived a surface gravity of log g $= 2.4$ based on the Gaia data\cite{Bailer_2018}. This log g is not inconsistent with the adopted value log g $= 2.7$. Radial velocity was measured from the line positions of the Fe absorption lines. The heliocentric radial velocity of J1124+4535 is 54.8 $\mpm$ $0.5$ km s$^{-1}$. Equivalent widths of absorption lines were measured by fitting Gaussian profiles. The elemental abundances were determined in the standard local thermodynamic equilibrium (LTE) manner, employing ATLAS model atmospheres with no convective overshooting\cite{Castelli_1997} and the MOOG code\cite{Sneden_1973}. The carbon abundance was derived by matching CH band at 4310 \AA. The spectrum-synthesis technique was used for the determination of Eu abundance (as shown in Supplementary Figure 1). The upper limits on the Er and Th abundances are estimated from Er II 4142 \AA~line and Th II 5989 \AA~line. A line list compiled from literature sources\cite{Roederer_2018,Ivans_2006} was adopted for abundance analysis. The hyperfine splitting is includes in the abundance determination for Sc, Mn and Ba based on the Kurucz database. The isotope ratios from the r-process component of the solar system\cite{Ishigaki_2013} are adopted for Ba and Eu. Non-LTE corrections for Na I, Mg I and Fe I lines are calculated based on the INSPECT database\cite{Lind_2011}. The average non-LTE correction of $-0.41$ dex for Na I abundance is included in the value presented in Table 1. The average non-LTE corrections for Mg I and Fe I lines are 0.03 dex and 0.06 dex, respectively. Since these two corrections are smaller than the errors in the derived abundances, we adopted the LTE abundances for Mg I and Fe I.\\   \textbf{Kinematic and orbital parameters.} The Galactic velocity components (U,V,W), the pericentric radius (r$_{\scriptsize \textmd{peri}}$), apocentric radius (r$_{\scriptsize \textmd{apo}}$), maximum height above the Galactic midplane (Z$_{\scriptsize \textmd{max}}$) and eccentricity (e) of the orbit of J1124+4535 are calculated from measurements of the heliocentric radial velocity, proper motion and distance\cite{Xing_2018}. The proper motion and distance are adopted from literatures based on Gaia DR2\cite{Lindegren_2018,Bailer_2018}. Adopting MWPotential2014 from galpy\cite{Bovy_2015} as the gravitational potential of the Milky Way, the orbital parameters of this object are estimated to be e $=0.58$, r$_{\scriptsize \textmd{peri}}$ $=3.46$ kpc, r$_{\scriptsize \textmd{apo}}$ $=12.85$ kpc, and Z$_{\scriptsize \textmd{max}}$ $=12.4$ kpc. Supplementary Figure 2 shows the Toomre diagram for J1124+4535, thick-disk and halo stars\cite{Xing_2018}. The kinematics of J1124+4535 is not inconsistent with the hypothesis that this star was accreted from a satellite dwarf galaxy, but the uncertainty is too large to derive any clear conclusion.

\end{methods}

\newpage
%additional references

\newpage

%% Use \item's to separate, default label is "Acknowledgements"

\begin{addendum}
 \item[Data availability]
The data that support the plots within this paper and other findings of this study are available from the corresponding author upon reasonable request.
%----------------------------------------------------------------------------------------
%Correspondence
\item[Correspondence]
Correspondence and requests for materials should be addressed to Gang Zhao (email: gzhao@nao.cas.cn) and Qian-Fan Xing (email: qfxing@nao.cas.cn).
%----------------------------------------------------------------------------------------
%   Acknowledgements
 \item[Acknowledgements] Support for this work was provided by the National Natural Science
 Foundation of China No.11890694, 11390371, 11603033, 11573032 and 11773033.
 This research was supported by JSPS-CAS Joint Research Program. W.A was supported by JSPS
 KAKENHI Grant Numbers 16H02168. M.N.I. was supported by JSPS KAKENHI Grant Number 17K14249.
 This paper includes data collected at the Subaru Telescope, which is operated by the National
 Astronomical Observatory of Japan. Funding for LAMOST (www.lamost.org) has
 been provided by the Chinese NDRC. LAMOST is operated and managed by the National Astronomical
  Observatories, CAS.
  %----------------------------------------------------------------------------------------
 %Competing interests
  \item[Competing interests]
The authors declare no competing interests.
  %----------------------------------------------------------------------------------------
%   Author Contributions
 \item[Author Contributions]
Q.F.X. takes responsible for the data analysis and paper draft . G.Z. proposed and initiated this research subject. W.A. and T.M. obtained the high-resolution spectra.?H.N.L. ,S.H. and M.N.I. assist Q.F.X for data reduction and interpretation of observed chemical abundances. The manuscript are further revised by Q.F.X., G.Z., and W.A. with all authors contributing comments and suggestions.

\end{addendum}

%\pagenumbering{gobble}

%\subimport{/Users/louise/Work/Papers/Nature15}{tables.tex}

\end{document}